\documentstyle[prl,aps,twocolumn,epsfig]{revtex}
\begin{document}
\draft
\title{
On a possible estimation of the constituent--quark
parameters from Jefferson Lab experiments
on pion form factor}

\author{A.F.Krutov\thanks{Electronic address:
krutov@info.ssu.samara.ru}}
\address {\it Samara State University, 443011, Samara, Russia}

\author{V.E.Troitsky\thanks{Electronic address:
troitsky@theory.npi.msu.su}}
\address {\it Nuclear Physics
Institute, Moscow State University, 119899, Moscow, Russia}

\date{August 1998}
\maketitle

\begin{abstract}
 The charge form factor of the pion is calculated for
momentum transfer range of Jefferson Lab experiments.
The approach is based on the instant form of the
relativistic Hamiltonian dynamics. It is shown
that the form-factor dependence on the choice
of the model for quark wave function in pion is weak,
while the dependence on the constituent--quark mass
is rather significant. It is possible
to estimate the mass of constituent quark and the sum of
anomalous magnetic moments of $u$- and $\bar d$- quarks from the
JLab experiments.
 \end{abstract}
\pacs{PACS numbers:\  12.39, 12.60.R, 13.40.G}

\narrowtext
At present time the constituent quark model (CQM) is widely
and successfully used for the description of hadron properties
at low and intermediate energies
\cite{GoI85,AzT80,ChC88,%
ItB90,JaK90,Jau91,KrT93,KiW94,Sch94,CaG96,Kru97,%
KrT98,BaK96}.
The reasons for this are well known:  first, CQM uses the
physically adequate degrees of freedom;
second, CQM describes nonperturbative effects. These
facts give a possibility to use CQM for the investigation
of the so called "soft" structure of hadrons, e.g. in
exclusive processes, in contrast to QCD (see, e.g.
\cite{IsL89}).

The main feature of CQM {\it versus} QCD is the extraction  of
finite number of the most important degrees of freedom needed to
describe the hadron. All dynamical effects of QCD are
incorporated in CQM through the effective (constituent) quark
mass and internal quark structure
in terms of quark form factors.
So, in the framework of CQM constituent quarks have all the
material properties of free particles and interact with each
other through the confinement potential. This means
that constituent quark is characterized
by an effective mass, a mean--square--radius (MSR) and an
anomalous magnetic moment.
Let us remark that
the concept of extended constituent  quarks also appears
in some quantum field theory models, for example, in Nambu--Jona--Lasinio
model with spontaneous chiral symmetry breaking \cite{VoL90}.
In this context
one can imagine that CQM is initiated by QCD.
However, it is very important to remind ourselves that CQM is not
a direct consequence of QCD, but a very successful
phenomenological model \cite{God97}.

For the description of electroweak properties it is
necessary to take into account the relativistic effects,
especially large in systems of light quarks.
We will use
the relativistic Hamiltonian dynamics (RHD) \cite{KeP91},
which is one of approaches to describe relativistic properties
of CQM.

In the present paper we discuss the dependence of
electromagnetic pion form factor on the internal quark structure.
The interest to this problem is due particularly to a
possible interpretation of  current experiments in Jefferson Lab
on the measurement of pion form factor \cite{CEBAF93} in the
range of momentum transfer $0.5 (GeV/c)^2\><\>Q^2\><\> 5(GeV/c)^2$ .
Using one of relativistic forms of CQM we obtain
that pion form factor in this region of $Q^2$ depends
strongly on the constituent quark mass, while the dependence
on model quark interaction in pion is weak. This fact gives hope
that it could be possible to estimate the
constituent quark mass from Jefferson Lab experiments.
With the use of model independent
Gerasimov sum rule \cite{Ger95} it is possible to estimate
the anomalous magnetic moments of constituent quarks
from these experiments and our calculations. So, the important
characteristics of CQM can be obtained.

In this paper we use the version
\cite{BaK95} of the instant form of RHD. In this approach we are
dealing with the following integral form of
pion electromagnetic form factor in the relativistic impulse
approximation:
\begin{equation} F_\pi (Q^2)=\int d\sqrt s\
d\sqrt{s'}\ \varphi (k(s))\,g_0(s,Q^2,s')\, \varphi (k'(s')).
\label{ff}
\end{equation}
Here $s = 4(k^2+M^2)$, $g_0(s,Q^2,s')$ is  the so called free
two--particle form factor to be derived by the methods
of relativistic kinematics \cite{BaK95},
$\varphi(k)$ is a phenomenological wave function
normalized with the account  of
relativistic density of states \cite{BaK95}:
\begin{equation}
\int\, \frac{dk}{2\,\sqrt{k^2 + M^2}}\,\varphi^2(k) =
1\>,\label{norm wf} \end{equation}

\begin{equation}
\varphi(k) = \sqrt[4]{4(k^2 + M^2)}\,k\,u(k),~~~
\int\,dk\,k^2\,u^2(k) = 1\>.\label{norm nrwf}
\end{equation}
\noindent
Here $u(k)$ is a nonrelativistic wave function.
The Eq.(\ref{norm wf}) gives the normalization of the
pion charge form factor $F_\pi(0)$ = 1.
The free two--particle form factor in (\ref{ff})
is of the form \cite{BaK95}:
$$
g_0(s,Q^2,s') = a(s,Q^2,s')
[\theta(s'-s_1)-\theta(s'-s_2)]
$$
$$
 \times \left\{b(s,Q^2,s' )[G^u_E(Q^2)+G^{\bar d}
_E(Q^2)]\cos \omega (s,Q^2,s')  \right.
$$
\begin{equation}
\left. + c(s,Q^2,s')
[G^u_M(Q^2)+G^{\bar d}_M(Q^2)]
\sin \omega (s,Q^2,s')  \right\}\>,
\label{g_0}
\end{equation}
where
$G^{u,\bar d}_{E}(Q^2)$ and $G^{u,\bar d}_{M}(Q^2)$ are
the electric and magnetic single-quark Sachs form factors,
$\omega (s,Q^2,s') $  is the Wigner rotation parameter.
The coefficients $a(s,Q^2,s'), b(s,Q^2,s'), c(s,Q^2,s')$
and $\omega (s,Q^2,s')$ and the value $s_{1,2}$ entering
the step function $\theta(s'-s_{1,2})$  are given by Eq.(10)
of Ref.~\cite{BaK96} (with $M_{\bar s} = M_{u} = M $).

By analogy with \cite{CaG96} and with the scaling of
nucleon form factors we write:
\begin{equation}
G^{q}_{E}(Q^2) =
e_q\,f(Q^2)\>,\quad G^{q}_{M}(Q^2) = (e_q +
\kappa_q)\,f(Q^2)\>,
\label{q ff}
\end{equation}
where $e_q$ is the quark charge and $\kappa_q$ is the quark
anomalous magnetic moment.
However, we do not use for $f_q(Q^2)$ the form of
Ref.\cite{CaG96} but that of Ref.\cite{Kru97}:
\begin{equation}
f_q(Q^2) =
\frac{1}{1 + \ln(1+ \langle r^2_q\rangle Q^2/6)}\>.
\label{f_qour}
\end{equation}
Here $\langle r^2_q\rangle$ is the quark MSR.
Let us discuss in brief the motivation for choosing the
explicit form (\ref{f_qour}). One of the features of our
approach is the fact that the form factor asymptotic behavior at
$Q^2\to\infty\>,\quad M\to$ 0
does not depend on the choice of the wave
function in (\ref{ff}) and is defined by the relativistic
kinematics of two--quark system only ~\cite{KrT98}.
In the point--like quark  approximation ($\kappa_q$=0,
$\langle r^2_q\rangle$= 0)  the asymptotics coincides with that
described by quark counting laws \cite{MaM73}:
$F_\pi(Q^2) \sim Q^{-2}$.  The
form (\ref{f_qour}) gives logarithmic corrections to the
power--law asymptotics , obtained in QCD. So, in our approach the form
(\ref{f_qour}) for the quark form factor gives the same
asymptotics as in QCD. The monopole form for $f_q(Q^2)$
\cite{CaG96} gives the different one.  Let us notice, however,
that the main results of the present paper do not depend on the
actual form of the quark form factor.

As the quark interaction potential is not known from  the
first principles, CQM is usually dealing with model potentials
and wave functions depending on  fitting parameters.  To
calculate pion form factor we use the following wave
functions for the ground state of quark--antiquark system:

1. Harmonic oscillator (HO) wave function (see e.g.
\cite{ChC88}):
\begin{equation} u(k)= N_{HO}\,
\hbox{exp}\left(-{k^2}/{2\,b^2}\right),
\label{HO-wf}
\end{equation}

2. Power-law (PL) wave function (see e.g. \cite{Sch94}):
\begin{equation}
u(k) =N_{PL}\,{(k^2/b^2 +
1)^{-3}}\>.  \label{PL-wf}
\end{equation}

3. The wave function with linear confinement and Coulomb--like
behavior at small distances \cite{Tez91}:
\begin{equation}
u(r) = N_T \,\exp(-\alpha r^{3/2} - \beta r)\>,
\label{Tez91-wf}
\end{equation}
$$
\quad \alpha =
\frac{2}{3}\sqrt{2\,M_r\,a}\>,\quad \beta = M_r\,b\>.
$$
In Eq.(\ref{Tez91-wf}), $a\>$ and $b$ are the parameters of
linear and Coulomb parts of potential respectively. We use the
value $b$=0.7867.

Let us emphasize that
we have obtained the expression (\ref{ff}) for the form factor
in the framework of the essentially relativistic approach:
instant form of RHD. Our current matrix element is explicitly
Lorentz covariant and satisfies conservation laws, so that the
current operator of composite system does contain the
contribution not only of one--particle currents but of
two--particle currents, too. We do not use fixed ("good")
current components or fixed coordinate frame (for example,
Breit frame), as one usually do in other RHD approaches
\cite{KeP91}.

One can see from the equations
(\ref{ff})--(\ref{Tez91-wf})
that we use the standard CQM parameters:
the constituent quark mass $M_u = M_d = M$,
the $u-$ and $\bar d-$ quark anomalous magnetic moments
$\kappa_u\>,\>\kappa_{\bar d}$
(which enter our equations through the sum
$s_q = \kappa_u + \kappa_{\bar d}$),
the constituent quark MSR
$\langle r^2_u\rangle = \langle r^2_d\rangle = \langle
r^2_q\rangle$ and the wave functions parameters -- $b$ in the
models (\ref{HO-wf}), (\ref{PL-wf}) and $a$ and $b$ in the model
(\ref{Tez91-wf}).

Let us notice that the electroweak properties of mesons
have been discussed by different authors
in the framework of CQM in the point--quark approximation
($\langle r^2_q\rangle = 0\>,\>\kappa_q = 0$)
and a consistent description of some processes
has been obtained \cite{ChC88,ItB90,Jau91,BaK96}.
However, there are strong arguments against this approximation.
The model independent Gerasimov sum
rules \cite{Ger95} indicate the existence of
anomalous magnetic moments of constituent quarks.
The anomalous
magnetic moments of quarks appear in the calculations
of Refs.\cite{AzT80,CaK96}.
In our approach the necessity to make the calculations of
electromagnetic and weak processes consistent
brings one in natural way to the concept of
the constituent--quark structure.

The parameters in our calculations are of two types. The first
type parameters enter the electromagnetic or weak current of
constituent quark: $M\>,\>s_q\>,\>\langle r^2_q\rangle$. The
second type parameters characterize quark interaction (wave
functions) -- $b\>,\>a$. We suppose that in the framework of CQM
the parameters of the first type do not depend on the parameters
of the second type: the first type parameters are the same for
different model interactions and these parameters are to be
fixed independently from the choice of model interaction.  In
other words, the calculation of composite quark systems is
analogous to that of composite nuclear systems, e.g.  the
deuteron. In calculation of the deuteron electromagnetic
properties one fixes the parameters in nucleon form factors
independently of the choice of nucleon--nucleon interaction
potential.

Let us now fix the parameters.
At the present time there are two pion characteristics that
can be extracted from the data
in a model independent way and with sufficient accuracy:
the mean square radius $ \langle r_\pi^2\rangle ^{1/2}_{exp} =
0.657\pm 0.012\> $ fm ~\cite{Ame84},  and the lepton decay
constant $ f_{\pi\>exp} = 0.1317\pm 0.0002\> $ GeV \cite{PDG}.
We assume that the calculations for any quark interaction
model satisfy (in addition to the description of the
particle spectrum) the conditions:
\begin{equation}
\langle r_\pi^2\rangle ^{1/2} =
\langle r_\pi^2\rangle ^{1/2}_{exp} \label{r}
\end{equation}
\begin{equation}
f_\pi = f_{\pi\>exp}.  \label{fpi}
\end{equation}
We have used the following forms for
pion MSR and the lepton decay constant
~\cite{BaK96}:
\begin{equation}
\langle r_\pi^2\rangle
=-\,6\,\left.\frac{dF_\pi(Q^2)}{dQ^2}\right|_{Q^2=0} =
\langle r^2_{r.m.}\rangle + \langle r^2_q\rangle .
\label{rpi} \end{equation} \begin{equation} f_\pi
=\frac{M\,\sqrt{n_c}}{\pi}\,\int\,\frac{k^2\,dk}{(k^2 + M^2)^{3/4}}\,
u(k)\>.  \label{dec}
\end{equation}
$n_c$ -- number of colors. In Eq.(\ref{rpi})
$\langle r^2_{r.m.}\rangle $ is the
contribution of quarks relative motion, it depends on
$M\>,\>s_q$ and on the wave functions parameters;
$\langle r^2_q\rangle $ is
the part of pion MSR due to MSR of quarks.  The lepton decay
constant is defined by the wave function parameters and by the
mass of constituent quark.

The choice of the values (\ref{r}), (\ref{fpi})
to fix the parameters has the following reasons.
First, they are measured in a model independent way, that is with
no assumptions about the pion structure.
Second, as one can see from the Eq.(\ref{rpi}),
the mean square radius of pion is determined by the
form factor behavior near zero. This means that
\begin{table*}
\caption{The values of model parameters for higher
($M$ = 0.22 GeV), medium  ($M$ = 0.25 GeV) and lower
($M$ = 0.33 GeV) groups of curves in Fig 1. The parameter $b$ in
(\protect\ref{HO-wf}),
(\protect\ref{PL-wf})
is in GeV, the parameter $a$ in (\protect\ref{Tez91-wf}) is in
GeV$^2$.  The wave functions parameters $b$, $a$ and the sum
$s_q$ of quark anomalous magnetic moments are derived
from the fitting of the pion MSR
$\langle r^2_\pi\rangle ^{1/2}$=0.657$\pm$0.012 fm
\protect\cite{Ame84} and the best possible posterior fitting of
the value $f_{\pi\>exp}$ = 131.7$\pm $0.2 MeV
\protect\cite{PDG}.  } \begin{center}
\begin{tabular}{|c|c|c|c|c|c|c|}
\hline
               &\multicolumn{2}{|c|}{$M$ = 0.22}
               &\multicolumn{2}{|c|}{$M$ = 0.25}
               &\multicolumn{2}{|c|}{$M$ = 0.33}\\
               &\multicolumn{2}{|c|}{$s_q$ = 0.0268}
               &\multicolumn{2}{|c|}{$s_q$ =-0.0023}
               &\multicolumn{2}{|c|}{$s_q$ =-0.1965}\\
\hline
Model               &$b, a$  &$f_\pi$&$b, a$   &$f_\pi$&$b, a$ &$f_\pi$\\
\hline
(\ref{HO-wf})       &0.3500  &127.4  &0.3069 &127.8&0.2558&125.1\\
\hline
(\ref{PL-wf})       &0.6131 &131.7 &0.5401&131.7&0.4901&131.7\\
\hline
(\ref{Tez91-wf})    &0.1331  &131.7 &0.0670 &132.1&0.0187&131.7\\
\hline
\end{tabular}
\end{center}
\end{table*}
\noindent
 the
condition (\ref{r}) gives, in fact, a constraint
for the pion form factor at small $Q^2$ values.
Analogously the constant $f_{\pi}$ is connected with the
pion form factor behavior at large momentum transfer.
So, the conditions (\ref{r}) and (\ref{fpi})
constrain, in fact, the pion form factor behavior at small
and large momentum transfer.
Let us note that in the point--quark approximation
the conditions ~(\ref{r}), ~(\ref{fpi}) can not be
fulfilled for realistic values of parameters
simultaneously for all models (see below).

So, the constituent quark parameters
$M\>,s_q,\>\langle r^2_q\rangle $ are the same for all the
wave functions
(\ref{HO-wf}), (\ref{PL-wf}), (\ref{Tez91-wf}).
The quark mass is a fitting parameter.
In addition, we shall use the relation
 $\langle r^2_q\rangle \simeq 0.3 /M^2$
between the MSR and the mass of the constituent quark
\cite{CaG96,VoL90}.

Let us consider now the parameter $s_q $ and the parameters of
the wave functions. To fix these parameters in the framework of
the model under consideration one can use the conditions
~(\ref{r}), ~(\ref{fpi}). This means however
that the values of $s_q $ are different
for different interaction models.
If we use one and the same value of $s_q $ for all the
models (\ref{HO-wf}), (\ref{PL-wf}), (\ref{Tez91-wf}),
we can not satisfy the conditions ~(\ref{r}), ~(\ref{fpi})
simultaneously. So, we shall use the parameters of wave
functions
(\ref{HO-wf}), (\ref{PL-wf}), (\ref{Tez91-wf}) and the parameter
$s_q$ as to fulfill the condition
~(\ref{r}) accurately
for all models. The condition ~(\ref{fpi})
will be satisfied only approximately. However, we try to
use the best possible value of $f_\pi$ for each model.
These values are given in the Table I. In such a way we fix all
but one parameters: the constituent quark mass  $M$ remains to
be a fitting parameter.

When one changes $M$ the other parameters are changed
following the indicated prescription,  $\langle r^2_q\rangle $,
$s_q$, $a$ and $b$ being the functions of $M$.

The results of our calculation of the pion form factor using the
parameters from the Table 1 indicate that the form factor
dependence on the quark interaction model is weak, while the
dependence on the constituent quark mass is rather significant.
Our results are presented in Fig.1. The curves calculated
with different wave functions but one and the same quark mass
form groups. The position of the group changes essentially
with the quark mass.

The great accuracy of planned JLab experiments will make it
possible to fix the position of "the group" rather accurately
and, so, to determine the constituent quark mass
\cite{note}.
This estimate (almost) will not  depend on the interaction model
for quarks in pion.

It is worth to emphasize that the function $f_q (Q^2)$
~(\ref{f_qour}) enters $F_{\pi} (Q^ 2)$ as a multiplier
and so the choice of $f_q (Q^2)$  does not influence the
relative position of curves for different $M$ and different
model wave functions.

It is possible that the slope of the experimental curve
will occur to be greater than in our groups, so that
for different $Q^2$ the points will belong to different groups.
This case will indicate the constituent quark mass depends
on the momentum transfer, in the spirit of the
Ref.\cite{KiW94}.
\begin{figure*}
\begin{center}
\epsfig{file=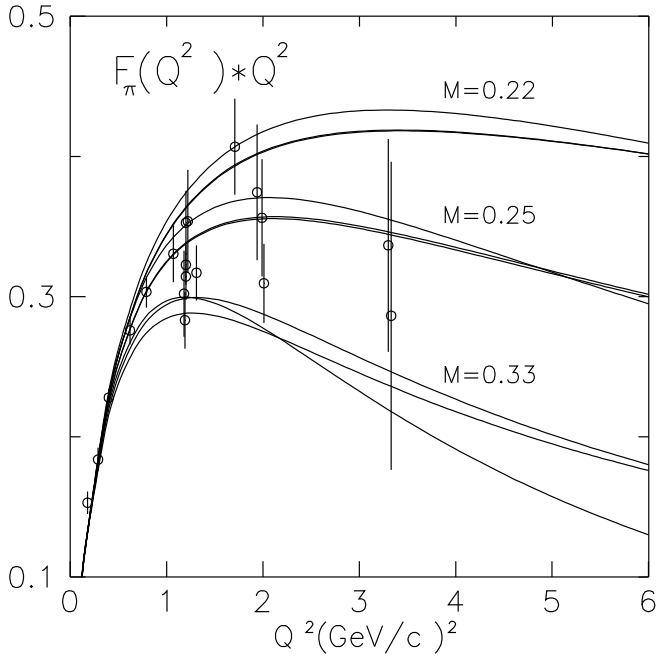,%
}
\end{center}
\caption{$\pi$--meson form factor in the range of JLab
experiments. The results of calculations for different
interaction models and $M$=0.22, 0.25, 0.33 GeV.
The curves with the same mass form a group.
The position of a group is defined by the constituent quark
mass.}
\end{figure*}

So, our approach gives the possibility to estimate the
constituent quark mass from experimental pion form factor.
Moreover, if the mass is determined
we can estimate the anomalous magnetic moments of
$u-$ and $d-$ quarks using the parameter
$s_q = \kappa_u + \kappa_{\bar d}$.
To perform this estimation one can use the model independent
Gerasimov sum rule \cite{Ger95}:
\begin{equation}
\frac{e_u + \kappa_u}{e_d + \kappa_d} =
-1.77\>,
\label{sumrul}
\end{equation}

For example, for $M \simeq 0.25$ GeV we obtain from
~(\ref{sumrul}):
$\kappa_u = -0.0285\>,\>\kappa_d =-0.0262$,
these values are of the order of the values of Ref.\cite{Ger95}.
The variation of $M$ gives different value of
$s_q$, and thus, of $\kappa_u $ and $\kappa_d$.

The analogous program can be carried out for the kaon.

To conclude, the calculation of the pion form factor in the
framework of our approach based on the instant form RHD gives
weak dependence on the interaction model for quarks in pion,
while the dependence on constituent quark mass is strong.
One can imagine
that any approach to the calculation of
the form factor with any wave function will give the
result close to our result, if the MSR and the lepton decay
constant are described well enough.
Our results provide a possibility to estimate the parameters
of the constituent quarks from JLab experiments.

The authors thank S.B.Gerasimov and N.P.Zotov for
helpful discussions and valuable comments. This work was
supported in part by The Russian Foundation for Basic Research
(grant No 96--02--17288).


\begin{references}
\bibitem{note}
Let us note that the use of our approach to the existing
experimental data (see Fig.1) gives only rough
estimate for the mass: 0.22$\><\>M\><$0.33 GeV.
\end{references}
\end{document}